\documentclass[preprint,floatfix]{revtex4-1}
\usepackage{amsmath,amssymb,amsfonts,mathrsfs,bbm,braket,xcolor,graphicx,graphics,latexsym,placeins,epsfig,multirow,makecell}
\usepackage{comment}
\usepackage{graphicx}
\usepackage{epsfig}
\usepackage{epstopdf}
\usepackage{amsfonts}
\usepackage{amssymb}
\usepackage{amsbsy}
\usepackage{amsmath}
\usepackage{mathrsfs}
\usepackage{latexsym}
\usepackage{natbib}
\usepackage{bm}
\usepackage{color}
\usepackage{braket}
\usepackage{slashed}
\usepackage{pgfplots}
\usepackage{graphicx,subcaption,lipsum}
\usepackage[normalem]{ulem}

\DeclareMathOperator{\sech}{sech}

\usepackage{tikz}
\usetikzlibrary{shapes,arrows,shadows}
\usepackage{hyperref,url}
\hypersetup{
	colorlinks=true,       
	linkcolor=blue,          
	citecolor=blue,        
	filecolor=blue,      
	urlcolor=blue           
}

\usepackage{enumitem}
\newlist{steps}{enumerate}{1}
\setlist[steps, 1]{label = Step \arabic*:}

\begin{document}

\affiliation{ Department of Physics, 
Indian Institute of Technology Bombay,
Mumbai - 400076, India }


\date{\today}

\title{Quantum signatures in black hole accretion: Pair production in dynamical magnetic fields}
%
%
\author{Susobhan Mandal}
\email{sm12ms085@gmail.com}
\affiliation{Department of Physics, Indian Institute of Technology Bombay, Mumbai 400076, India}
\author{Tausif Parvez}
\email{214120002@iitb.ac.in}
\affiliation{Department of Physics, Indian Institute of Technology Bombay, Mumbai 400076, India}
\author{S. Shankaranarayanan}
\email{shanki@iitb.ac.in}
\affiliation{Department of Physics, Indian Institute of Technology Bombay, Mumbai 400076, India}	%

\begin{abstract}
Accretion disks around black holes host extreme conditions where general relativity and magnetohydrodynamics dominate. 
These disks exhibit two distinct dynamical regimes --- Standard and Normal Evolution (SANE) and Magnetically Arrested Disk (MAD). In the MAD regime, these systems exhibit magnetic fields up to \(10^8\) G and variability on gravitational timescales \(t_g \sim 10^{-4}\) s for stellar-mass black holes. While classical magnetohydrodynamics has been extensively applied, quantum effects in these high-energy environments remain unexplored. 
Here, we employ quantum field theory in background gauge fields (QFTBGF) to demonstrate that the dynamic magnetic fields of MADs drive significant pair production via the Schwinger mechanism. The resulting pairs emit \emph{non-thermal (synchrotron) radiation}  with a peak frequency tunable across $1–3000$ MHz, depending on the magnetic field strength (peaking at higher frequencies for stronger fields). For $B \sim 10^8$ G,  our model predicts a peak spectral flux density of $\sim 1–100$ mJy, detectable with next-generation radio telescopes (e.g., SKA, ngVLA). This work provides a direct and observable signatures of quantum effects in black hole accretion disks.
\end{abstract}

\maketitle

\section{Introduction}
Accretion disks, spanning systems from stellar-mass black holes (StMBHs), like Cygnus X-1, to supermassive black holes  (SuMBHs), like  M87*, exhibit extreme physical conditions: gas temperatures of $10^7 – 10^{12}$ K, magnetic fields of $10^3 – 10^8$ G, and velocities approaching relativistic speeds ($v/c \sim 0.1–0.3$)~\cite{2006csxs.book..381F,Bosch-Ramon:2011afb,Abramowicz:2011xu,Abramowicz:2014,2019PhyU...62.1126S}. These environments are governed by general relativity (GR), magnetohydrodynamics, and radiative processes~\cite{Gammie:2003qi,Davis:2020wea,Vigano:2020ouc}, producing observable signatures across the electromagnetic spectrum --- from radio (GHz–THz) to gamma rays (MeV–TeV)~\cite{Sironi:2015eoa,Yuan:2014gma,Matthews:2020lig}. Disk emission often shows variability on timescales of milliseconds (for StMBHs) to years (for SuMBHs), while relativistic jets can extend up to kiloparsecs, with luminosities reaching $10^{47}$ erg/s in blazars~\cite{Brandenburg:1995abc,Beckwith:2008abc,Tchekhovskoy:2011zx,EventHorizonTelescope:2019dse,EventHorizonTelescope:2019pcy}. 

Accretion flows exhibit distinct dynamical regimes governed by the 
plasma-$\beta$ parameter, defined as the ratio of plasma pressure $P_\text{gas}$ 
to magnetic pressure $P_\text{mag} = B^2/8\pi$~\cite{Chandrasekhar:1961ab,Jafari:2019brc}. This parameter critically determines the flow’s stability and structure~\cite{Balbus:1998ab}. For instance, in Standard and Normal Evolution (SANE)~\cite{Narayan:2012yp,Blandford:2018iot,Hovatta:2019ulp,Matthews:2020lig,Begelman:2021ufo} disks, $\beta \sim 1 - 10$, while in the case of Magnetically Arrested Disk (MAD),  
$\beta \ll 1$ (typically \(10^{-3}\)–\(10^{-1}\) near the horizon)~\cite{Tchekhovskoy:2011zx}. In SANE disks, the accretion process is primarily governed by turbulence, resulting in a relatively disordered magnetic field structure. In contrast, MADs are characterized by a strong, coherent magnetic field that exerts significant pressure on the infalling matter to arrest the the accretion flow close to the BH horizon~\cite{Begelman:2021ufo}. 
In MAD, strong poloidal fields (\(B \sim 10^4\)–\(10^8\) G)~\cite{EventHorizonTelescope:2021bee} halt inward accretion~\cite{Ichimaru:1977uf}, powering jets via the Blandford-Znajek mechanism~\cite{Blandford:1977ds,Lee:1999se,McKinney:2005zw}.
However, direct observational evidence conclusively confirming the MAD conjecture has remained elusive~\cite{Beckwith:2008abc,Liska:2018btr}. This is due, in part, to the fact that the predicted signatures of MADs are often obscured by other processes occurring within the accretion flow (like Disk winds/outflows)~\cite{Yuan:2015vfa}, and that current observational techniques are limited by the resolution~\cite{EventHorizonTelescope:2019dse} and frequency coverage~\cite{Weltman:2018zrl}.

In this work, we propose a novel approach to probe the 
quantum effects leading to particle creation driven by the dynamical magnetic fields. The timescale for a MAD to form around a BH is estimated to be $t_{\rm MAD} \sim 10^{4} \, r_g/c$, corresponding to roughly $10^{5}\,{s}$ ($1\,{s}$) for SuMBH (StMBH)~\cite{Narayan:2021qfw,Pathak:2025duv}. This rapid formation timescale underscores the dynamic nature of MADs and the potential for significant \emph{large-scale} time-dependent electromagnetic phenomena. Specifically, we apply quantum field theory in background gauge fields (QFTBGF) techniques~\cite{Birrell:1982ix,Fulling:1989nb,Mukhanov:2007zz,Parker:2009uva,Hu:2020luk} to analyze the particle production in the strong, time-varying magnetic fields of a MAD. This time variation  induces electric fields, leading to a dynamically evolving electromagnetic environment~\cite{Hawley:2015qma}. Since, $\beta \ll 1$ near the horizon, as we show, the particle production scales with $B$ and its variability $(\partial_t B \propto \sqrt{P_{\rm mag}})$. 

\begin{figure*}[ht]
\begin{minipage}[b]{0.45\linewidth}
\centering
\vspace*{-0.5cm}
\includegraphics[width=\textwidth]{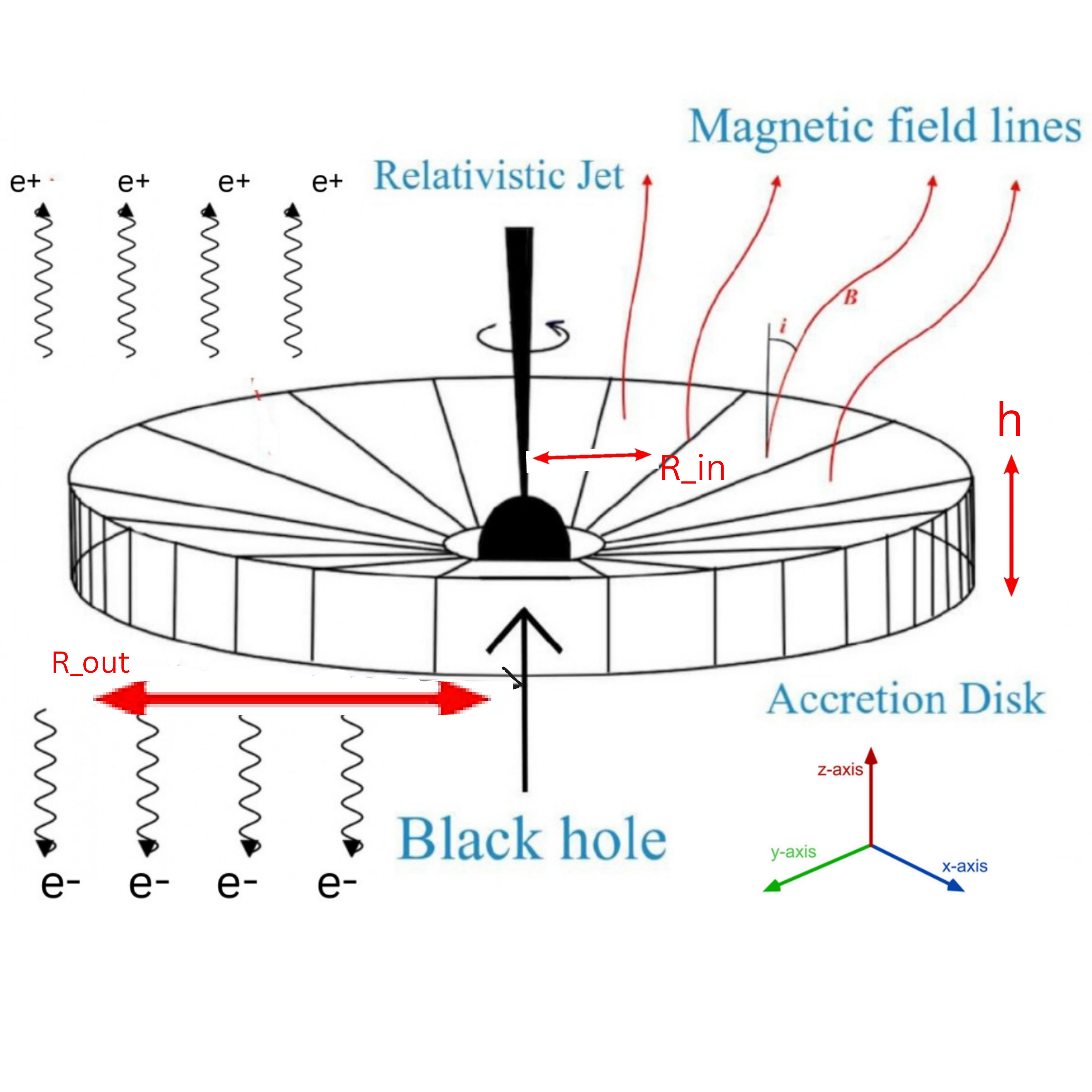}
\caption{Evolving magnetic field in the accretion disk leads to particle production}
\label{fig:figure1}
\end{minipage}
\hspace{0.5cm}
\begin{minipage}[b]{0.45\linewidth}
\centering
\includegraphics[width=0.8\textwidth]{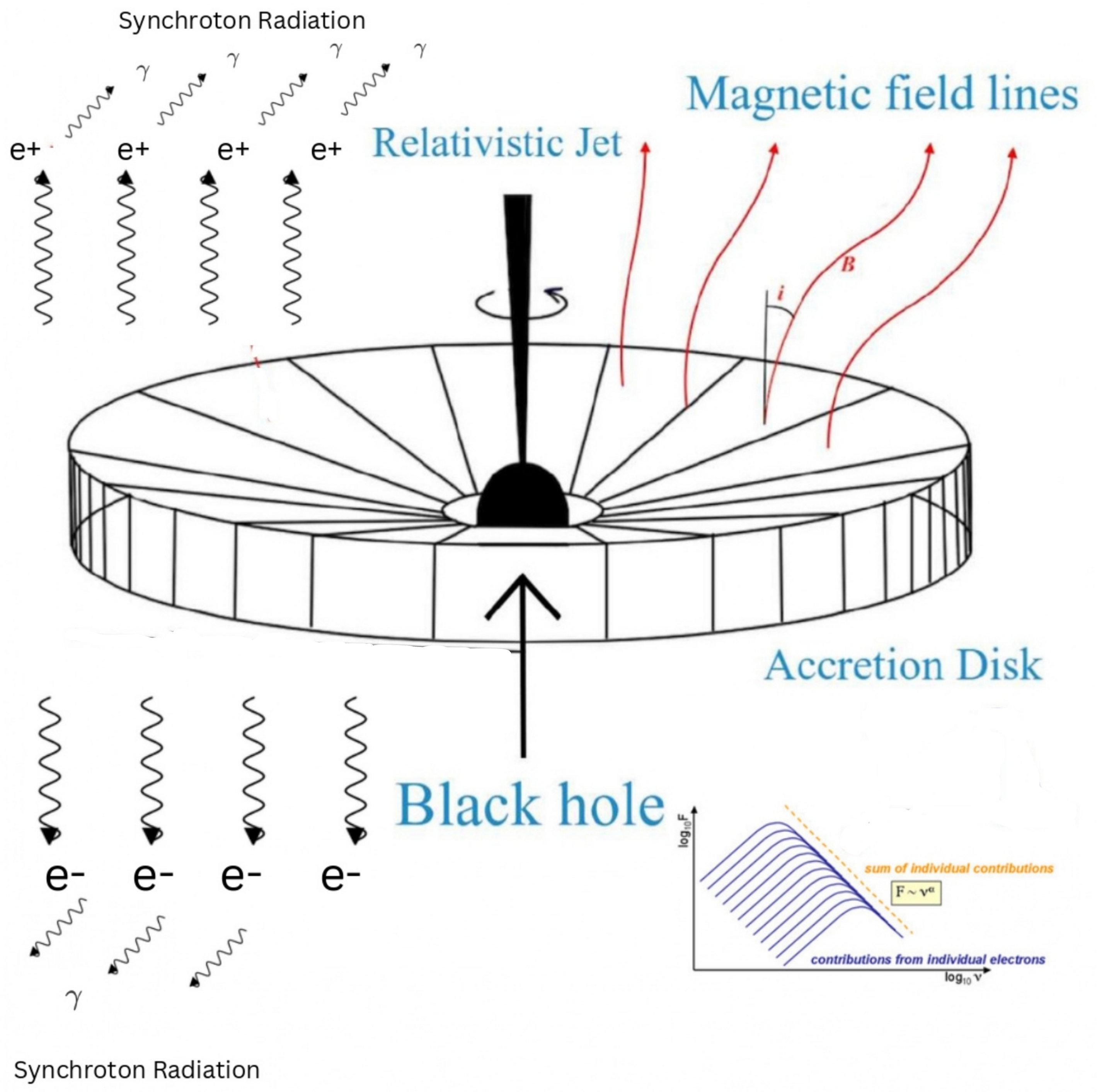}
\caption{Produced charged particles accelerate due to the magnetic field 
leading to synchrotron radiation. (Illustration from Ref.~\cite{Jafari:2019brc})}
\label{fig:figure2}
\end{minipage}
\end{figure*}

While the gravitational field strength is relatively weak at the distances where accretion disks form around StMBH or SuMBH~\cite{2022-Shanki.Joseph-GRG}, the very existence of these disks, and the MAD phenomena they exhibit, are fundamentally linked to the presence and properties of BH~\cite{Hawley:2015qma}.  
The produced pairs are accelerated due to the presence of large-scale, ordered magnetic fields inherent to the MAD configuration, leading to the emission of synchrotron radiation~\cite{Rybicki:847173,Merloni:2003aq}. This radiation, emitted by relativistic charged particles spiraling along magnetic field lines, is often beamed in the direction of their motion due to relativistic effects, enhancing its observed intensity. The resulting pairs emit synchrotron radiation with a peak frequency tunable across $1 - 3000$ MHz, with stronger magnetic fields correlating with higher peak frequencies. For 
magnetic fields in the range $10^7\,G$ to  $10^8\,G$, our model predicts a peak spectral flux density of $1-100\, {\rm mJy}$, potentially detectable with next-generation radio telescopes such as the Square Kilometer Array (SKA) and next-generation Very Large Array (ngVLA)~\cite{SKA,nGVLA,DSA-2000,Haarlem:2013,FASTVLBI}. 

\section{The setup}
The central BHs in these systems are well-modeled by the Kerr spacetime, which captures their spin dynamics~\cite{EHT2019I}. {However, since the accretion disk typically forms at radii 
\(\gtrsim 10^{3} - 10^{4}~r_{g}\)~\cite{Kemp:1980ab,EHT2019IV,Narayan:1997ku,Murchikova_2019}, the local spacetime curvature --- quantified by the Kretschmann scalar \(\mathcal{K} \sim r_g^2/r^6\) --- becomes negligible (\(\mathcal{K}^{1/2} \ll 1 \, \text{cm}^{-2}\)) for both StMBH and SuMBH~\cite{MTW,2022-Shanki.Joseph-GRG}.} To isolate the dominant effects of time-dependent magnetic-field driven particle production, we approximate the disk’s local environment as \emph{Minkowski spacetime} with an external magnetic field $B \sim 10^4\)–\(10^8$ G~\cite{Tchekhovskoy:2011zx}. To avoid gauge ambiguities~\cite{Parker:2009uva}, we model the disk’s matter as a \emph{charged complex scalar field} $\Phi$  coupled to the external electromagnetic field $\mathcal{A}_\mu$, governed by the action:  
\begin{equation}\label{action}
\mathcal{S} = \int d^{4}x \ \Big[(\mathcal{D}_{\mu}\Phi)^{\dagger}\mathcal{D}^{\mu}\Phi - m^{2}
\Phi^{\dagger}\Phi\Big],
\end{equation}
where $m$ is the mass of $\Phi$ (taken to be electron mass) and $\mathcal{D}_{\mu} = \partial_{\mu} - ie\mathcal{A}_{\mu}$, $\mathcal{A}_\mu$ satisfies Lorentz gauge condition $\partial_{\mu}\mathcal{A}^{\mu} = 0$.  We set $c = \hbar = 1$, $e = \sqrt{4 \pi \alpha_e}$, $\alpha_e = 1/137$~\cite{Kolb:1990vq} and the metric signature $(+, -, -, -)$.

Observations of SuMBHs reveal strong, poloidal magnetic fields in their inner regions, believed to be responsible for launching powerful jets~\cite{Abramowicz:2011xu}. These fields are oriented vertically near the black hole and arc over the accretion disk. In our local Minkowski approximation for the accretion disk environment, we will consider an external magnetic field that reflects the strength and general characteristics of these observed poloidal fields, allowing us to investigate its quantum effects on matter fields. This approach provides a tractable framework for studying QFTBGF~\cite{Greiner:1985ce} in MADs.

In this setup,  Fig. \eqref{fig:figure1}, the rotational motion of charged plasma particles due to angular momentum in the accretion disk generates a dominant magnetic field component along the z-direction. While a purely azimuthal current would imply an electric field perpendicular to the $x-y$ plane of the disk, the presence of a $U(1)$ charged current with a non-zero z-component introduces an electric field component parallel to the z-axis ($E_{\parallel}$). To model accretion disk with azimuthal symmetry, evolving towards a MAD state, we consider a time-dependent vector potential in the \emph{symmetric gauge} (chosen to describe a z-directed magnetic field and a parallel component):
\begin{equation}
\mathcal{A}_{\mu} = (0, - B(t)y/2, B(t)x/2, - A_{\parallel}(t))
\label{eq:GaugeChoice}
\end{equation}
where $B(t)$ and $A_{\parallel}(t)$ are time-dependent functions. 
%
We model the time-varying potentials as representing a transient behavior associated with MAD dynamics:
\begin{equation}
\label{eq:At-Bt}
A_{\parallel}(t) = a_{\parallel}\sech\left( t/\Delta \right) \, , \,
B(t) = B_{0}\exp \left(- {t^{2}}/{\Delta^{2}} \right)
\end{equation}
where $\Delta$ is the typical timescale ($t_{\rm MAD}$) of the magnetic field evolution to the MAD state both in StMBH and SuMBH~\cite{DelSanto:2013ab,EventHorizonTelescope:2019pcy}, and $a_{\parallel}, B_0$ are constants related to the magnetic field strength in the MAD state. The electric field along the z-axis is $E_{\parallel} \sim a_{\parallel}/\Delta$. $B(t)$ is sustained by the current associated with the flow in the plasma inside the accretion disk. However, it is not confined to the accretion disk, as the total flux associated with $B(t)$ in a closed surface has to be zero. $t_{\rm MAD}$ takes into account the accretion rate and BH spin.
To model StMBH/SuMBH~\cite{DelSanto:2013ab,EventHorizonTelescope:2019pcy}, the constants are taken to be:
\begin{equation}
\label{eq:constants}
a_{\parallel}/m = 10^{-12},  \, (m \, \Delta)^{-1} = 10^{-27}, \, B_0 = 10^{6}~{\rm G} \, .
\end{equation}
Hence, $E_{\parallel}/B_{0} \sim 10^{-30}$ implying magnetic dominance.

Given the setup, our analysis proceeds in two key steps to evaluate quantum signatures in black hole accretion disks and compare them with observations:
\begin{steps}
\item {\bf Quantum Particle Production during Magnetic Field Evolution:} 
We compute the pair production via the Schwinger mechanism driven by the time-dependent electromagnetic fields during the SANE-to-MAD transition (see Fig.~\eqref{fig:figure1}). The particle production rate is quantified by Bogoliubov coefficients~\cite{Birrell:1982ix,Parker:2009uva,Dabrowski:2016tsx} which encode mode mixing induced by the evolving fields $B(t)$ and $A_\parallel(t)$. While our results hold for general adiabatic field profiles, we adopt specific forms given in Eq.~\eqref{eq:At-Bt}. Even a weak 
electric field $E_\parallel$ accelerates the pairs along the jet axis, seeding relativistic outflows.

\item {\bf Synchrotron radiation from produced particles:} 
The pairs gyrate in strong poloidal fields  ($10^4 < B_0 ({\rm G}) < 10^8$)  emitting synchrotron radiation \cite{Rybicki:847173} (See Fig.~\eqref{fig:figure2}). We derive the non-thermal spectral energy distribution across a frequency range of $1 - 3000$ MHz, providing a direct link between the prediction and observations.
\end{steps}

\section{Particle production during magnetic field evolution} 

To go about this, we begin with the equation of motion for the charged complex scalar field $\Phi$, derived from the action \eqref{action} in the symmetric gauge \eqref{eq:GaugeChoice}:
\begin{equation}
\label{EOM 1}
\Big[ \partial_{t}^{2} + \left[- i\nabla_{\perp} - e \mathbf{A}_{\perp}(t)\right]^{2} + [- i\partial_{z} - e A_{\parallel}(t)]^{2} + m^{2}\Big]\Phi = 0 \, ,
\end{equation}
where $\mathbf{A}_{\perp}(t) = (- B(t)y/2, B(t)x/2, 0)$. Introducing conjugate momentum operators~\cite{Wilczek:1989ab}: $\Pi_{x,y} = - i\nabla_{\perp} - e \mathbf{A}_{\perp}(t)$ and $\Pi_{z} = - i\partial_{z} - e A_{\parallel}(t)$, we utilize the ladder operators $\hat{\cal C} = (\Pi_{x} - i\Pi_{y})/{\omega_B(t)}$ and $\hat{\cal C}^{\dagger} = (\Pi_{x} + i\Pi_{y})/{\omega_B(t)}$, with $\omega_{B}^2(t) = 2|eB(t)|$, satisfying $[\hat{\cal C}, \hat{\cal C}^{\dagger}] = 1$ and $\Pi_{x}^{2} + \Pi_{y}^{2} = \omega_{B}^2(t)(\hat{\cal C}^{\dagger}\hat{\cal C} + 1/2)$. Using these definitions, Eq.~\eqref{EOM 1} can be expressed as:
\begin{equation}
\label{eq:EOM02}
\Big[\partial_{t}^{2} + \Pi_{\perp}^{2}(t) + \Pi_{z}^{2}(t) + m^{2}\Big]\Phi = 0.
\end{equation}
Since the symmetric gauge \eqref{eq:GaugeChoice} has no z-dependence and the above equation corresponds to  3-D time-dependent harmonic oscillator, we choose the following ansatz: 
\[
\Phi_{n, l, k_{\parallel}}(t, \vec{x}) 
= e^{ik_{\parallel}z}\mathcal{H}_{n,l}(x\omega_{B}/\sqrt{2},y\omega_{B}/\sqrt{2})f_{n, k_{\parallel}}(t) \, ,
\]
where $\mathcal{H}_{n,l}$ refers to bivariate Hermite polynomials~\cite{Dunkl:2014ab}. Substituting ansatz in \eqref{eq:EOM02} and setting $\tau = m \, t$, we get:
\begin{eqnarray}\label{exact mode eqns}
\Big[
\partial_{\tau}^{2} + \frac{\omega_{n, k_{\parallel}}^{2}}{m^2}
+ \frac{\hat{\mathcal{D}} \mathcal{H}_{n,n}}{ \mathcal{H}_{n,n}}  
\left[ \frac{1}{\omega_{B}}\frac{d^2 \omega_B }{d\tau^2} 
+ \frac{2}{\omega_{B}} \frac{d \omega_B }{d\tau} \partial_\tau  \right]  & &  \\
+ \left(\frac{1}{\omega_B}  \frac{d \omega_B }{d\tau} \right)^2  \frac{\hat{\mathcal{D}}^2 \mathcal{H}_{n,n}}{\mathcal{H}_{n,n}}  \Big] f_{n, k_{\parallel}} &=& 0,~~~\nonumber
\end{eqnarray}
where $\hat{\mathcal{D}} $ is the 2-D dilatation operator, which commutes with the angular momentum operator ($\hat{L}_z$) along z-axis.
\begin{eqnarray}
\hat{\mathcal{D}} & \equiv& x \, {\partial}/{\partial x} + y \, 
{\partial}/{\partial y}   \\
\frac{\omega_{n,k_{\parallel}}^{2}}{m^2} &\equiv& 
 1 + \frac{1}{m^2} [k_{\parallel} + eA_{\parallel}(t)]^{2} 
+ \frac{\omega_B^2(t)}{m^2} \left[n + \frac{1}{2}\right] . \nonumber 
\end{eqnarray}
$\omega_{n, k_{\parallel}}(t)$ does not depend on the $l$, hence, leading to a large degeneracy like in quantum hall effect~\cite{Wilczek:1989ab} and the degeneracy factor is included in Eq.~\eqref{def:Nt}.

Interestingly the above expression can be further simplified using the fact that the derivatives of $\omega_B(t)$ for typical accretion disk parameters
\eqref{eq:constants} are small, i. e. 
{\small
\begin{equation}
\frac{d\ln \omega_{B}}{d\tau}  \sim \frac{1}{(m \Delta)}
\sim  10^{- 27} \, , \, 
\frac{d^{2}\ln\omega_{B}}{d\tau^{2}} 
\sim \frac{1}{(m\Delta)^{2}} \sim 10^{- 54}. 
\end{equation}
}
Substituting in Eq.~(\ref{exact mode eqns}) yields the simplified equation (details in Appendix \eqref{AppendixA}):
\begin{equation}
\label{eq:C-THDO}
(\partial_{t}^{2} + \omega_{n, k_{\parallel}}^{2}(t))f_{n, k_{\parallel}} \simeq 0 \, ,
\end{equation}
which corresponds to the decoupled harmonic oscillators with time-dependent frequency $\omega_{n, k_{\parallel}}(t)$.
Decomposing the charged scalar field operator $\Phi_{k_{\parallel},n}(t)$ into Bosonic creation 
and annihilation operators for particles and anti-particles, we write
\begin{equation}\label{mode-decomposition}
\Phi_{k_{\parallel},n}(\tau) = a_{n,k_{\parallel}}f_{n,k_{\parallel}}(\tau) + b_{n, - k_{\parallel}}
^{\dagger}f_{n, - k_{\parallel}}^{*}(\tau),
\end{equation} 
where $f_{n,k_{\parallel}}(\tau)$ satisfies Eq.~\eqref{eq:C-THDO} and the Wronskian condition $f_{n,k_{\parallel}}^{*}\partial_{\tau}{f}_{n,k_{\parallel}} - \partial_\tau{f}_{n,k_{\parallel}}^{*}f_{n,k_{\parallel}} = - i$.

The time-dependent frequency $\omega_{n,k_{\parallel}}(t)$ in Eq.~(\ref{eq:C-THDO}) prevents a unique separation into particle- antiparticle pairs~\cite{Parker:2009uva}. However, for slowly varying fields, an adiabatic particle number can be defined using a reference basis of positive/negative energy plane waves~\cite{Birrell:1982ix,Parker:2009uva,Dabrowski:2016tsx}. This is achieved via a Bogoliubov transformation linking initial and final asymptotic vacuum states. This transformation defines time-dependent creation/annihilation operators, $\tilde{a}_{k}(t)$ and $\tilde{b}_{k}(t)$, related to the initial operators $a_{n,k_{\parallel}}$ and $b_{n,k_{\parallel}}$ from (\ref{mode-decomposition}) by:
\begin{equation}\label{BGT}
\begin{bmatrix}
\tilde{a}_{n,k_{\parallel}}(t)\\
\tilde{b}_{n, - k_{\parallel}}^{\dagger}(t)
\end{bmatrix} = \begin{bmatrix}
\alpha_{n,k_{\parallel}}(t) & \beta_{n,k_{\parallel}}^{*}(t)\\
\beta_{n,k_{\parallel}}(t) & \alpha_{n,k_{\parallel}}^{*}(t)
\end{bmatrix}
\begin{bmatrix}
a_{n,k_{\parallel}}\\
b_{n, - k_{\parallel}}^{\dagger}
\end{bmatrix} .
\end{equation}
where the matrix coefficients satisfy $|\alpha_{n,k_{\parallel}}(t)|^{2} - |\beta_{n,k_{\parallel}}(t)|^{2} = 1$ at all times. The time-dependent adiabatic particle number associated each $(n, k_{\parallel})$ quantum numbers are defined by the expectation value \textit{w.r.t} the initial vacuum state $\ket{0}$ (at $t \to -\infty$)~\cite{Dabrowski:2016tsx}. The total number of particles produced in the mode $(n, k_{\parallel})$ at late times  (at $t \to \infty$) is:
\begin{equation}\label{total number of particles}
\tilde{N}_{n,k_{\parallel}} 
= |\beta_{n,k_{\parallel}}(t \rightarrow \infty)|^{2}.
\end{equation}

We numerically {solved the time-evolution equations of $\alpha_{n,k_{\parallel}}, \beta_{n,k_{\parallel}}$ (given in Appendix \ref{AppendixA})} with the field profiles from Eq. (\ref{eq:At-Bt}) and initial conditions $\alpha_{n,k_{\parallel}}(-\infty) = 1, \beta_{n,k_{\parallel}}(-\infty) = 0$. The timescales for MAD formation, $t_{\rm MAD}$, were set to $\sim 10^{5} \, s$ for SuMBH and $1\, s$ for StMBH, consistent with astrophysical estimates~\cite{Narayan:2021qfw,Pathak:2025duv}.

\begin{figure}
\hspace*{-0.95cm}
\includegraphics[width = 14.0cm,height=10cm]{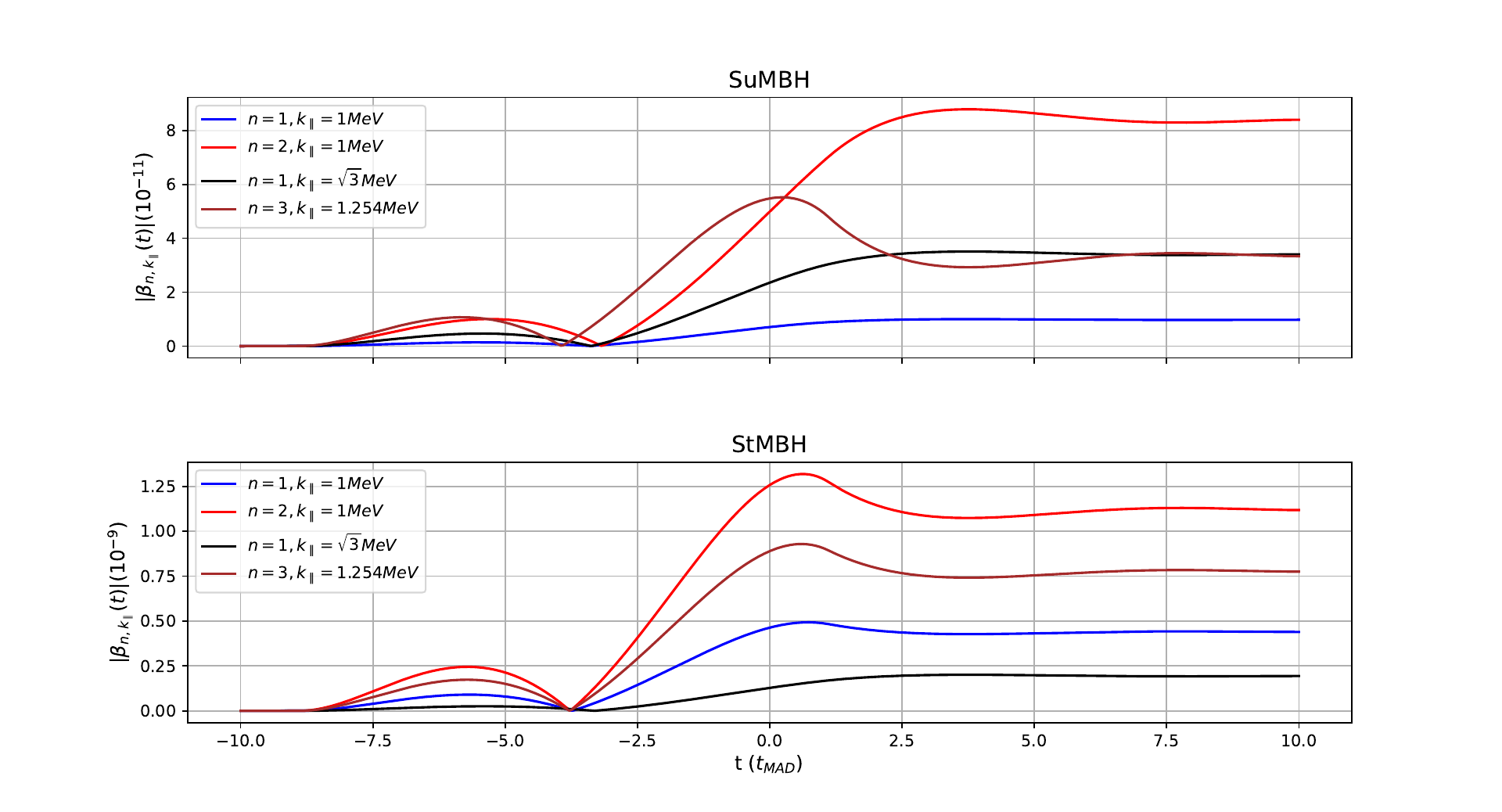}
\caption{Plot of $|\beta_{n,k_{\parallel}}|$ as a function of $t/t_{MAD}$. For SuMBH (StMBH), $B_0 = 10^4\,{\rm G} (10^5 \, {\rm G})$.}
\label{fig:figure3}
\end{figure}

Fig.~\ref{fig:figure3} illustrates the temporal evolution of $|\beta_{n, k_{\parallel}}|$ for representative parameter sets with $m = 0.5~{\rm MeV}$. The plots indicate that $|\beta_{n,k_{\parallel}}(t)|$ asymptotically saturates, suggesting sustained particle production even as the rate of field variation decreases. The magnitude of $|\beta_{n,k_{\parallel}}(t)|$ exhibits dependence on the Landau level $n$~\cite{Wilczek:1989ab}, but generally remains on the order of $10^{-11}$.

\begin{figure}
\hspace*{-0.95cm}
\includegraphics[width = 14.0cm,height=10cm]{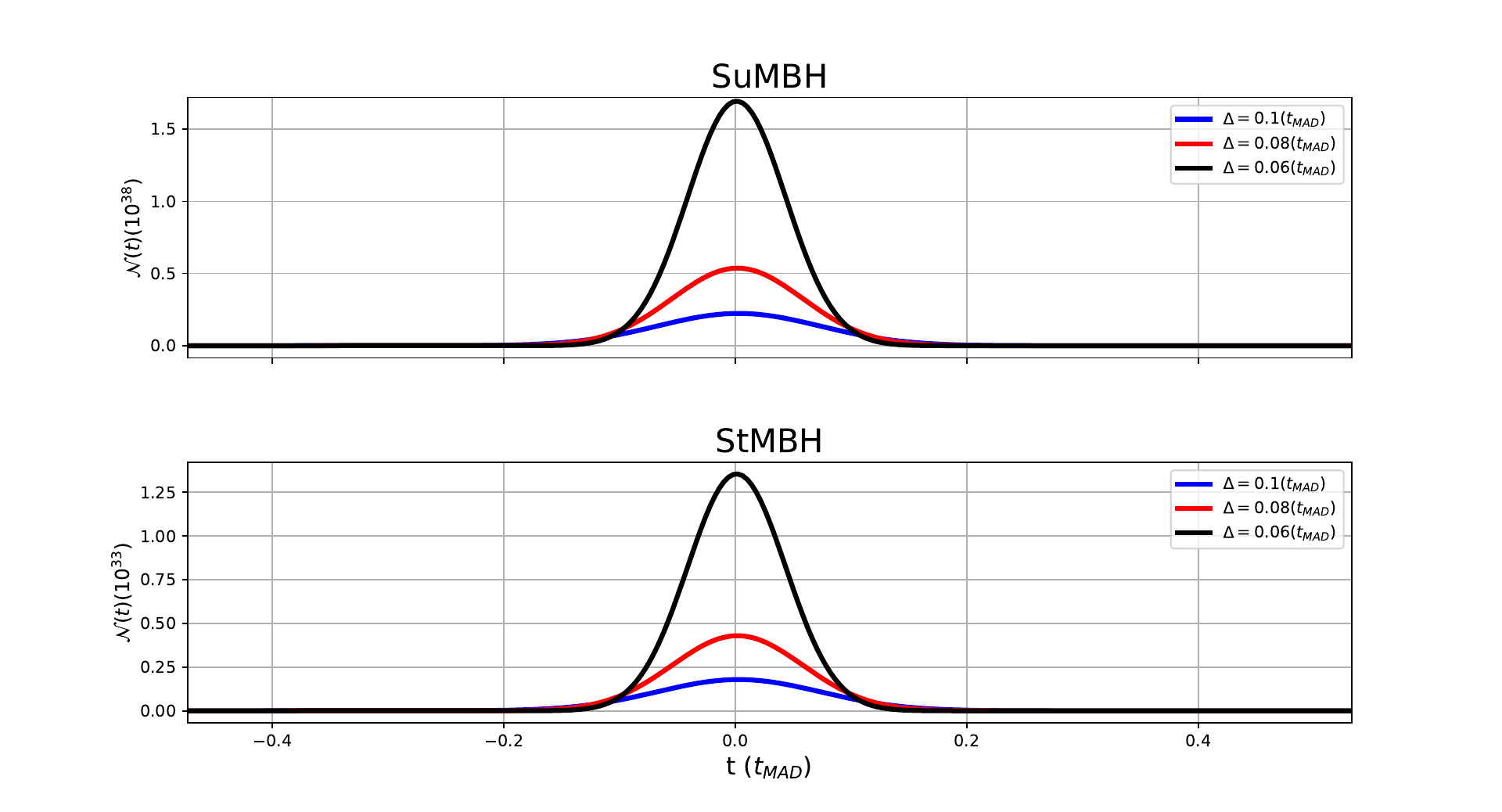}
\caption{Plot of number of particle-antiparticle pairs generated  $\mathcal{N}(t)$ during evolution. {We set $R_{\rm in} = 10^{3}r_{g}, R_{\rm out} = 10^{4}r_{g}, h = r_{g}$ \cite{Broderick:2008sp, Kemp:1980ab, Chauvin:2018ywd}.}} 
\label{fig:figure4}
\end{figure}

We now evaluate the number of particle-antiparticle pairs  generated during this evolution. 
The total number of particle-antiparticle pairs generated during the magnetic field evolution is given by the expectation value of the number operator integrated over all modes:
\begin{equation}
\mathcal{N}(t) = V\frac{|eB(t)|}{4\pi^{2}}\sum_{n}\int_{- \infty}^{\infty}dk_{\parallel} \ |\beta_{n, k_{\parallel}}|^{2} \, .
\label{def:Nt}
\end{equation}
where $V$ is the 3-D volume ($\pi (R_{\rm out}^2 - R_{\rm in}^2) \times h)$ (see Fig.~\ref{fig:figure1}) and $|eB(t)|/(4 \pi^2)$ is the degeneracy of the each $n$ mode. Fig.~\ref{fig:figure4} illustrates $\mathcal{N}(t)$ for different values of $\Delta$. While the specific number of produced particles depends on the timescale of field variation, the overall order of magnitude remains around $10^{38}$ (for SuMBH).

\section{Synchrotron radiation from produced particles:}

The charged particles produced during the SANE-to-MAD transition are accelerated to \emph{relativistic velocities} (with Lorentz factor $\gamma \sim 10^2$) by the MAD’s strong, ordered magnetic fields ($B_0 \sim 10^4\)–\(10^8$ G) emitting synchrotron radiation with a peak frequency~\cite{Rybicki:847173,Pabmanabhan:2000book}  
\[
\omega_\text{c} = 3\gamma^2  {eB}/({4\pi  m_e c)} \sim 1 \text{--}3000~\text{MHz} \, .  
\]  

The spectral flux density at a distance $d$ is~\cite{Rybicki:847173,Pabmanabhan:2000book} (details in Appendix \eqref{AppendixB}):
\begin{equation}
P_{\rm tot} \approx \frac{\sqrt{3} e^3 B}{4 \pi d^2 m_e c^2} \int_{0}^{\infty} \mathcal{N}(\mathcal{E}) F\left(\frac{\omega}{\omega_c},\mathcal{E}\right) d\mathcal{E},
\label{eq:Luminosity}
\end{equation}
where \(F(x) = x \int_x^\infty K_{5/3}(\xi) d\xi\) is the synchrotron kernel function, and \(\mathcal{N}(\mathcal{E})\) is the energy distribution of the produced particles, obtained via the Fourier transform of $\mathcal{N}(t)$~\footnote{$\mathcal{N}(t)$ is real, therefore, $\mathcal{N}^*({\cal E}) = \mathcal{N}^(-{\cal E})$. On the other hand, $\mathcal{N}(t) = \mathcal{N}(t)$ is an even function (by our choice of $B(t)$ and $A_{\parallel}(t)$); hence, $\mathcal{N}({\cal E}) = \mathcal{N}(-{\cal E})$. Combining these two things, we obtain, $\mathcal{N}({\cal E}) = \mathcal{N}^*({\cal E})$ and hence, $\mathcal{N}(\mathcal{E})$ is real.}. Following the integration of Eq.~\eqref{eq:Luminosity} to determine the spectral flux density as a function of $\omega$, we consider specific astrophysical sources. For StMBH, we adopt the distance to Cygnus X-1 ($d = 1.9 kpc$), while for SuMBH, we use the distance to Sgr A* ($d = 8.18 kpc$). 
\begin{figure}
\hspace*{-0.70cm}
\includegraphics[width = 14.0cm,height=10cm]{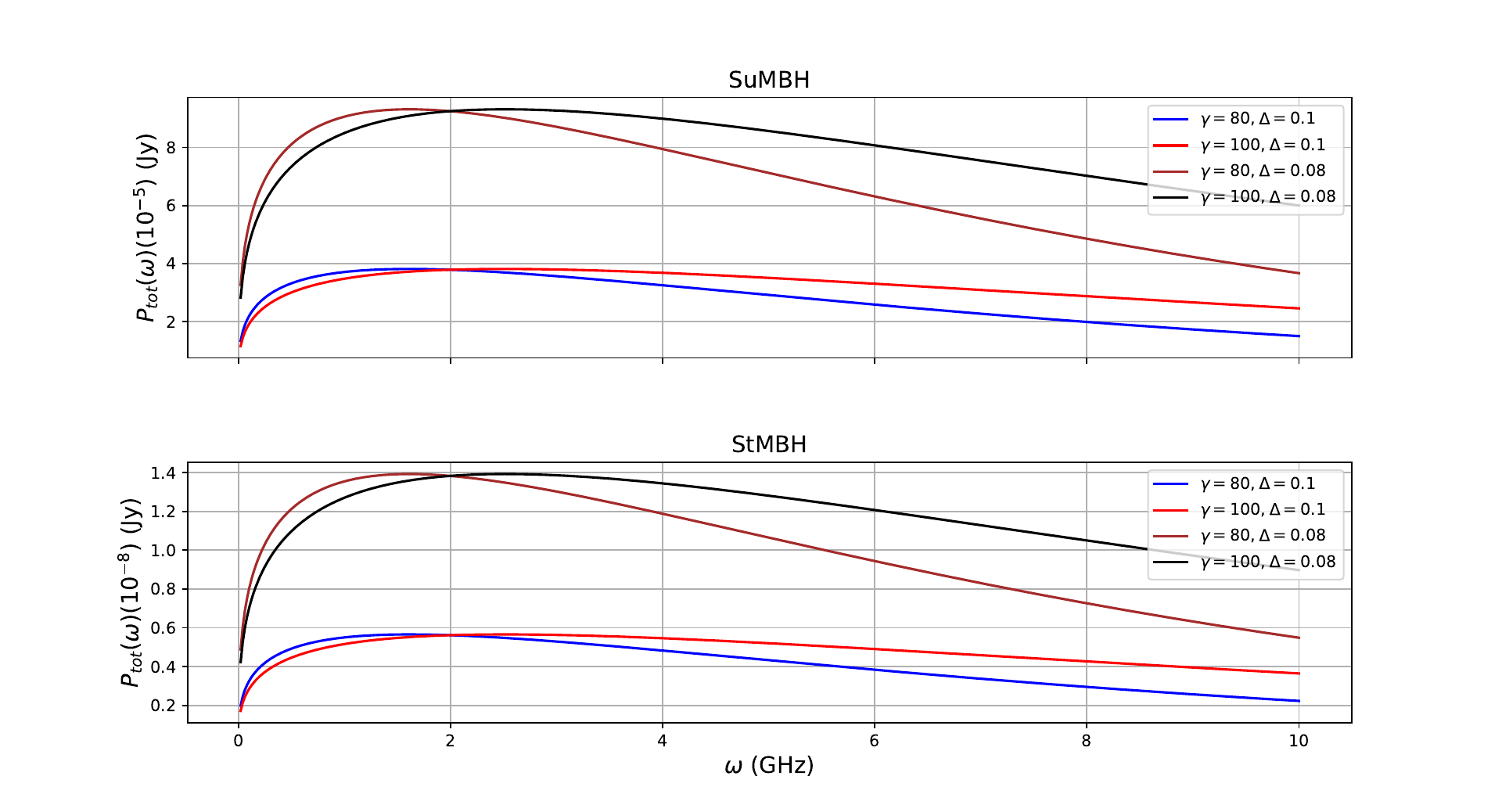}
\caption{Plot of $P_{\rm tot}(\omega)$ as a function $\omega$ for $B_0 = 10^7~{\rm G}$.}
\label{fig:figure5}
\end{figure}

Fig.~\ref{fig:figure5} presents the calculated spectral flux density for $B_0 = 10^7~{\rm G}$, revealing a peak emission frequency around $2~{\rm GHz}$ for both StMBH and SuMBH scenarios. The peak flux density exhibits dependence on the parameter $\Delta$ and a weaker sensitivity to the Lorentz factor $\gamma$ via $\omega_c$. The peak spectral flux density for a SuMBH is approximately $60 \mu{\rm Jy}$, while for a StMBH it is around $10~{\rm nJy}$.  These theoretical results suggest potential observational implications for next-generation radio telescopes.

\section{Observational Prospects with Next-Generation Radio Telescopes:}
This work presents the first detailed application of QFTBGF to investigate particle production within the dynamical magnetic environment of a MAD, establishing a novel theoretical connection between quantum processes and potential astrophysical observables. {Unlike classical synchrotron emission scenarios in BH jets, which primarily consider radiation from electrons, our quantum framework predicts radiation from both electron and positron pairs.} Consequently, synchrotron radiation will be emitted by both species. This is particularly significant for jet observability: even \emph{if a jet is primarily aligned away from Earth}, the synchrotron radiation from positrons produced within the accretion disk and potentially ejected at different angles could still be detectable. Our key prediction is a characteristic synchrotron emission across a frequency range of $1-3000$ MHz, with potentially detectable flux densities of $1$ to $100$ mJy for strong magnetic fields. 

These predictions from our work align remarkably well with the capabilities of the upcoming SKA and ngVLA~\cite{SKA,nGVLA,Corsi:2025tab}. Specifically, SKA-Low and SKA-Mid covering $350$~MHz to $15.4$~GHz,  comfortably encompasses our predicted 1-3000 MHz emission range, ensuring that our predicted emission falls within the observable window of these facilities.
Similarly, the ngVLA, planned to operate from 1.2 GHz to 116 GHz, also covers a portion of our predicted frequency window. Beyond SKA and ngVLA, facilities like DSA-2000 (operating at $0.7 - 2$~GHz)~\cite{DSA-2000} and LOFAR 2.0 (operating at $10 - 250$~ MHz)~\cite{Haarlem:2013} could independently validate detections, while FAST offers deep, targeted follow-up~\cite{FASTVLBI}. Our predicted flux densities of $1-100$~mJy , exceed the $\mu$ Jy sensitivity of these telescopes,  making detection feasible~\cite{Afonso:2001Ap,Whalen:2023abc,Willner:2023ApJ,Galluzzi:2025zwg}. This indicates that if the predicted synchrotron emission from electron-positron pairs in MADs exists at these levels, it should be well within the detection capabilities of these next-generation telescopes.

Detecting this characteristic radio emission would provide compelling observational support for the MAD paradigm and would offer direct empirical evidence for quantum processes occurring in the extreme gravitational environments around black holes. This unique signature, arising from quantum particle production, offers a new avenue to probe fundamental physics in the extreme conditions of black hole accretion disks, potentially revealing aspects of jet formation and composition that are inaccessible through purely classical treatments. 
Such observations would mark a significant step in probing fundamental physics in the universe's most extreme corners and underscore the pivotal role of future radio astronomy in advancing our understanding of quantum phenomena in astrophysical settings. The \emph{SKA, ngVLA and complementary facilities} offer unprecedented opportunities to observationally test the theoretical predictions and opens a new window into the interplay between quantum field theory and extreme gravity. {The distinct non-thermal frequency range differentiates this synchrotron emission from typical accretion disk radiation; however, potential contamination from the surrounding plasma requires further investigation.}

\noindent {\bf Acknowledgements}
The authors thank  I. Chakraborty, S. M. Chandran, P. G. Christopher,
A. Chowdhury, K. Hari, S. Jana, A.Kushwaha, B. Mukhopadhyay, M. Mukhopadhyay and  A. Naskar for comments and discussions. 
The authors thank SERB-CRG grant for funding this research. The MHRD fellowship at IIT Bombay financially supports TP.

\appendix 

\section{Particle production phenomenon in a time-dependent background}
\label{AppendixA}

To avoid gauge ambiguities~\cite{Parker:2009uva}, we model the disk’s matter as a \emph{charged complex scalar field} $\Phi$  coupled to the external electromagnetic field $\mathcal{A}_\mu$, governed by the action \eqref{action}.  To model the accretion disk evolving towards or exhibiting a MAD state, we consider a time-dependent vector potential in the \emph{symmetric gauge} \eqref{eq:GaugeChoice}. 

The equation of motion of the complex scalar field $\Phi$ of the above action in the above gauge leads to \eqref{EOM 1}. Choosing the following ansatz: 
\[
\Phi_{n, l, k_{\parallel}}(t, \vec{x}) 
= e^{ik_{\parallel}z}\mathcal{H}_{n,l}(x\omega_{B}/\sqrt{2},y\omega_{B}/\sqrt{2})f_{n, k_{\parallel}}(t) \, ,
\]
where $\mathcal{H}_{n,l}$ refers to bivariate Hermite polynomials~\cite{Dunkl:2014ab}, the mode-solutions $f_{n, k_{\parallel}}(t) $ satisfy the following equation, representing decoupled harmonic oscillators with a time-dependent frequency:
\begin{equation}\label{EOM 2}
(\partial_{t}^{2} + \omega_{n,k_{\parallel}}^{2}(t))f_{n,k_{\parallel}} = 0,
\end{equation}
where $\omega_{n,k_{\parallel}}^{2}(t) = \sqrt{2|eB(t)|\left(n + \frac{1}{2}\right) + (k_{\parallel}
+ eA_{\parallel}(t))^{2} + m^{2}}$.
 The time-dependent nature of the background electromagnetic sources breaks time-translation symmetry, leading to a time-dependent frequency $\omega_{n,k_{\parallel}}^{2}(t)$ that prevents a unique separation into particle-antiparticle pairs. However, for slowly varying fields, an adiabatic particle number can be defined using a reference basis of positive/negative energy plane waves, achievable via a Bogoliubov transformation. This demonstrates the phenomenon of particle production as a consequence of the time-dependent electromagnetic field.


Decomposing the charged scalar field operator $\Phi_{k_{\parallel},n}(t)$ into Bosonic creation
and annihilation operators for particles and anti-particles, \eqref{mode-decomposition}, we impose the familiar Wronskian relation
\begin{equation}\label{Wronskian relation}
f_{n,k_{\parallel}}^{*}\dot{f}_{n,k_{\parallel}} - \dot{f}_{n,k_{\parallel}}^{*}f_{n,k_{\parallel}}
= - i,
\end{equation}
on the solutions, which follows from the Klein-Gordon inner product definition.

The Bogoliubov transformation defines a set of time-dependent creation and annihilation operators,
$\tilde{a}_{k}(t)$ and $\tilde{b}_{k}(t)$, related to the original set of time-independent operators $a_{n,k_{\parallel}}$ and $b_{n,k_{\parallel}}$, appearing in (\ref{mode-decomposition}), by the following linear transformation \eqref{BGT}. For scalar fields, unitarity requires the matrix coefficients to satisfy $|\alpha_{n,k_{\parallel}}(t)|^{2} - |\beta_{n,k_{\parallel}}(t)|^{2} = 1$ for all time $t$. This transformation can be interpreted as rewriting the mode decomposition of the field operator in (\ref{mode-decomposition}) as:
\begin{equation}\label{mode-decomposition 2}
\begin{split}
\Phi_{k_{\parallel},n}(t) & = a_{n,k_{\parallel}}f_{n,k_{\parallel}}(t) + b_{n, - k_{\parallel}}
^{\dagger}f_{n, - k_{\parallel}}^{*}(t)\\
 & = \tilde{a}_{n,k_{\parallel}}(t)\tilde{f}_{n,k_{\parallel}}(t) + \tilde{b}_{n, - k_{\parallel}}^{\dagger}(t)\tilde{f}_{n, - k_{\parallel}}^{*}(t),
\end{split}
\end{equation}
where $\tilde{f}_{n, k_{\parallel}}(t)$ is a set of reference mode functions that reduce to free
particle solutions at the initial time when both the magnetic and electric fields vanish. Equivalently, the exact solutions $f_{n, k_{\parallel}}(t)$ to the oscillator equation (\ref{EOM 2}) can be decomposed in terms of these reference functions:
\begin{equation}\label{BGT 2}
f_{n,k_{\parallel}}(t) = \alpha_{n, k_{\parallel}}(t)\tilde{f}_{n,k_{\parallel}}(t) + \beta_{n,
k_{\parallel}}(t)\tilde{f}_{n, - k_{\parallel}}^{*}(t).
\end{equation}
The time-dependent adiabatic particle number associated with each $(n, k_{\parallel})$ quantum number is defined by the expectation value with respect to the original vacuum state $\ket{0}$ of the time-dependent number operators $\tilde{a}_{n,k_{\parallel}}^{\dagger}(t)\tilde{a}_{n,k_{\parallel}}(t)$:
\begin{equation}\label{adiabatic number}
\tilde{\mathcal{N}}_{n,k_{\parallel}}(t) \equiv \bra{0}\tilde{a}_{n,k_{\parallel}}^{\dagger}(t)
\tilde{a}_{n,k_{\parallel}}(t)\ket{0} = |\beta_{n,k_{\parallel}}(t)|^{2}.
\end{equation}
The total number of particles produced in the mode $(n, k_{\parallel})$ at late times is given by \eqref{total number of particles}. Our primary interest here is the full time evolution of the adiabatic particle number
$\tilde{\mathcal{N}}_{n,k_{\parallel}}(t)$, as it evolves from an initial value of zero to
its final asymptotic value $\tilde{N}_{n,k_{\parallel}} = \tilde{\mathcal{N}}_{n,k_{\parallel}}
(t \rightarrow \infty)$.

The Bogoliubov transformation, and hence the associated particle number $\tilde{\mathcal{N}}_{n,k_{\parallel}}(t)$ defined in (\ref{adiabatic number}), depends on the choice of the reference basis mode solutions, $\tilde{f}_{n,k_{\parallel}}(t)$. We choose these reference mode solutions to be of the following form:
\begin{equation}\label{reference mode solutions}
\tilde{f}_{n,k_{\parallel}}(t) \equiv \frac{1}{\sqrt{2W_{n,k_{\parallel}}(t)}}e^{-i\int^{t}
W_{n,k_{\parallel}}(t')dt'},
\end{equation}
where $W_{n,k_{\parallel}}(t)$ is a solution to the time-dependent oscillator equation (\ref{EOM 2}) and is related to $\omega_{n,k_{\parallel}}(t)$ as:
\begin{equation}\label{W-function}
W_{n,k_{\parallel}}^{2} = \omega_{n, k_{\parallel}}^{2} - \Big[\frac{\ddot{W}_{n,k_{\parallel}}}
{2W_{n,k_{\parallel}}} - \frac{3}{4}\left(\frac{\dot{W}_{n,k_{\parallel}}}{W_{n,k_{\parallel}}}\right)^{2}\Big].
\end{equation}
At the leading order of an adiabatic expansion, the solution to the above equation is $W_{n,k_{\parallel}}(t) = \omega_{n, k_{\parallel}}(t)$. In this case, the reference basis solutions in (\ref{reference mode solutions}) correspond to the standard WKB solutions of equation (\ref{EOM 2}).

Since the Bogoliubov transformation is a linear canonical transformation on the field operators, we also need to specify the associated transformation of the momentum field operator $\Pi_{n,
k_{\parallel}}(t) = \dot{\Phi}_{n,k_{\parallel}}^{\dagger}(t)$. Consequently, we must also
specify the decomposition of the first derivative of $f_{n,k_{\parallel}}(t)$ in terms of the
mode functions, and the general form consistent with unitarity is:
\begin{equation}
\begin{split}
\dot{f}_{n,k_{\parallel}}(t) & = \left(- iW_{n,k_{\parallel}}(t) + \frac{1}{2}V_{n,k_{\parallel}}
(t)\right)\alpha_{n,k_{\parallel}}(t)\tilde{f}_{n,k_{\parallel}}(t)\\
 & + \left(iW_{n,k_{\parallel}}(t) + \frac{1}{2}V_{n,k_{\parallel}}(t)\right)\beta_{n,k_{\parallel}}
(t)\tilde{f}_{n, - k_{\parallel}}^{*}(t).
\end{split}
\end{equation}
\begin{figure}[!htb]
\includegraphics[height = 10cm, width = 14cm]{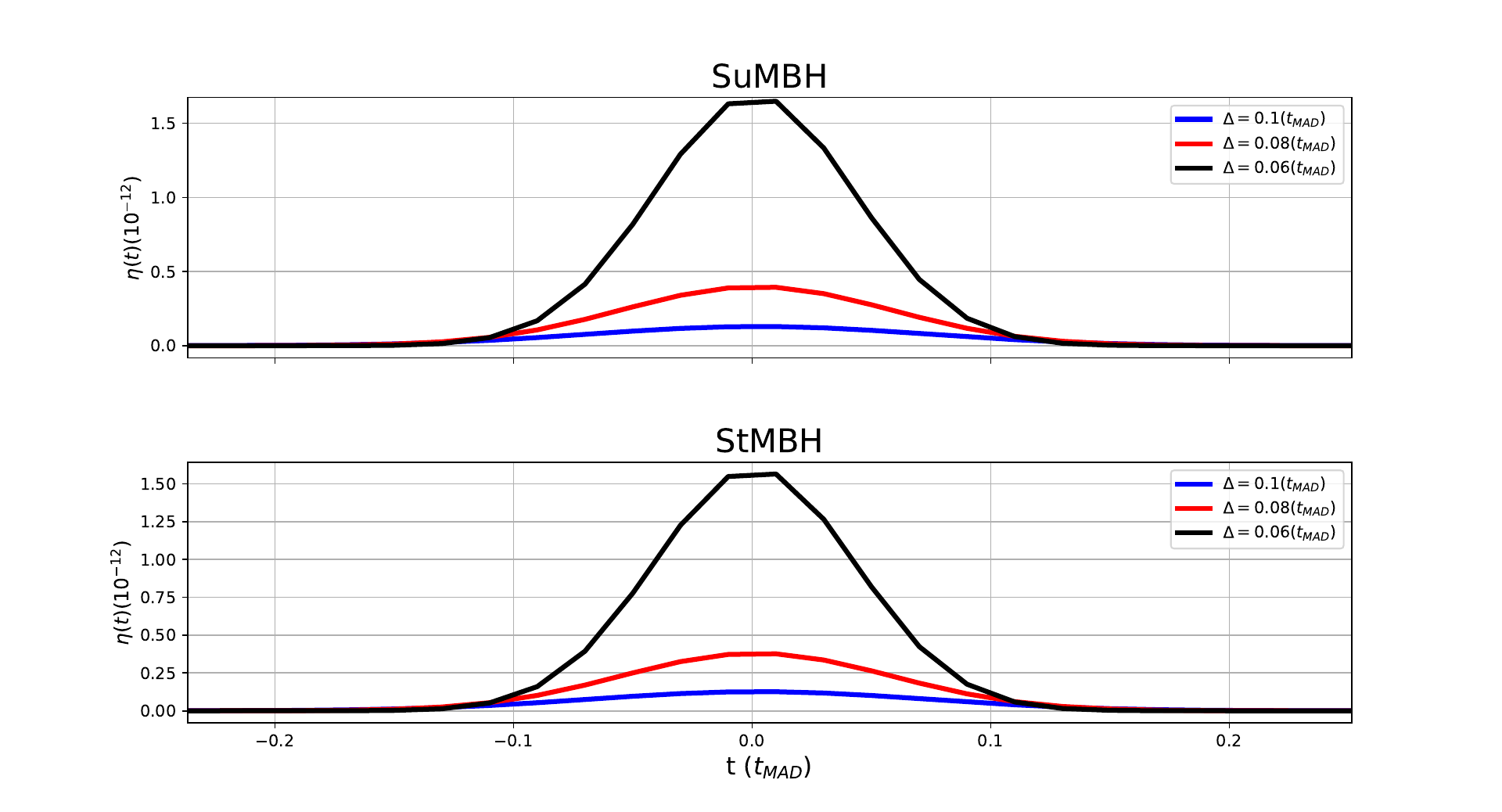}
\caption{Plot of particle production efficiency in the cases of SuMBH and StMBH with the same value of $B_{0}$.}
\label{Figure 1}
\end{figure}  

\begin{figure}
\begin{center}
\includegraphics[height = 10cm, width = 14cm]{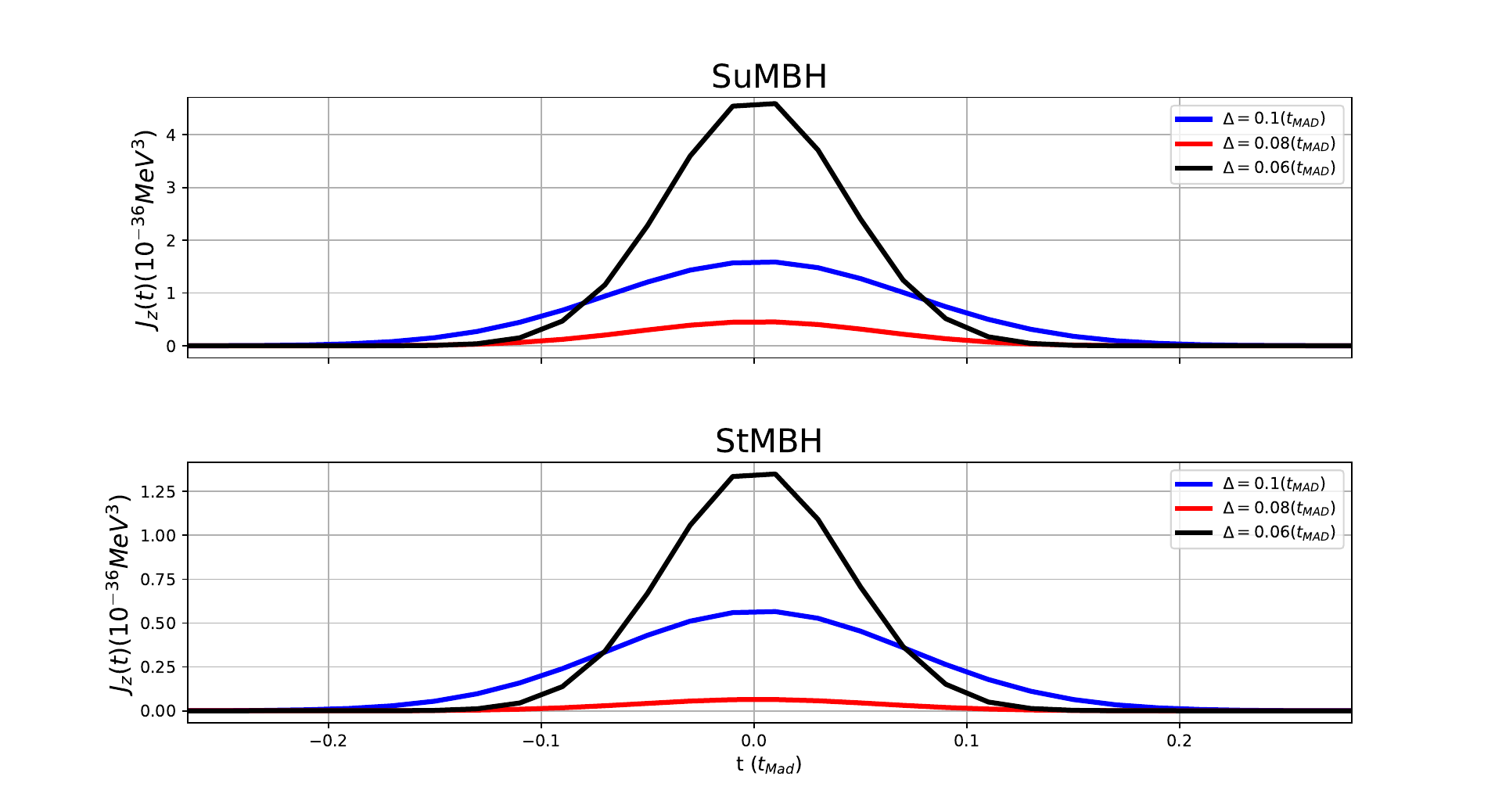}
\end{center}
\caption{Plot of gauge invariant $U(1)$ current in z-direction in the cases of SuMBH and StMBH due to the particle production with the same value of $B_{0}$.}
\label{Figure 2}
\end{figure}

In the above equations, $V_{n,k_{\parallel}}(t)$ is chosen such that the identification of positive and negative energy solutions can be made at the initial time. However, the freedom in the choice of $W_{n,k_{\parallel}}(t)$ and $V_{n,k_{\parallel}}(t)$ leads to some arbitrariness in defining these solutions at intermediate times. Here, we consider a conventional choice based on the WKB approximation, taking $W_{n,k_{\parallel}}(t) = \omega_{n,k_{\parallel}}(t)$ and $V_{n,k_{\parallel}}(t) = 0$. Considering this choice, we finally obtain the following set of coupled differential equations:
\begin{equation}
\begin{bmatrix}
\dot{\alpha}_{n,k_{\parallel}}(t)\\
\dot{\beta}_{n,k_{\parallel}}(t)
\end{bmatrix} = \frac{\dot{\omega}_{n,k_{\parallel}}(t)}{2\omega_{n,k_{\parallel}}(t)}  
\begin{bmatrix}
0 & e^{2i\int^{t}\omega_{n,k_{\parallel}}(t')dt'}\\
e^{ - 2i\int^{t}\omega_{n,k_{\parallel}}(t')dt'} & 0
\end{bmatrix}
\begin{bmatrix}
\alpha_{n,k_{\parallel}}(t)\\
\beta_{n,k_{\parallel}}(t)
\end{bmatrix},
\end{equation}
subject to the initial conditions $\alpha_{n,k_{\parallel}}(t \rightarrow - \infty) = 
1, \ \beta_{n,k_{\parallel}}(t \rightarrow - \infty) = 0$~\cite{Birrell:1982ix, Dabrowski:2016tsx, Parker:2009uva}. Once we know the solution of $\beta_{n,k_{\parallel}}(t)$, we can use it to compute the gauge-invariant and conserved $U(1)$ 4-current defined by
\begin{equation}
J^{\mu} = ie\left(\Phi^{*}\mathcal{D}_{\mu}\Phi - (\mathcal{D}_{\mu}\Phi)^{*}\Phi\right) \, . 
\end{equation}
From the above expressions, we can now, in principle, compute the efficiency of particle production, $\eta(t)$, in the time-varying magnetic field as:
\begin{equation}
\eta(t) = \frac{|eB(t)|}{4\pi^{2}}\sum_{n}\int_{-\infty}^{\infty}dk_{\parallel}\frac{2m_{e}}
{\omega_{n,k_{\parallel}}(t)}|\beta_{n, k_{\parallel}}(t)|^{2} \times \frac{1}{\frac{1}{2T}\int_{-T/2}
^{T/2}\langle\vec{E}^{2}(t') + \vec{B}^{2}(t')\rangle dt'},
\end{equation}
where $\frac{1}{T}\int_{-T/2}^{T/2}dt'$ represents the time average over a sufficiently large interval $T$, which for our numerical solution is chosen to be $T = t_{\rm MAD}$, and $\langle . \rangle$ denotes the spatial average. Fig. \ref{Figure 1} illustrates the temporal evolution of $\eta(t)$ for a fixed magnetic field amplitude ($10^{5}$ Gauss) and varying values of $\Delta$, the timescale associated with the changes in the magnetic and electric fields. Similarly, Fig. \ref{Figure 2} displays the temporal evolution of the z-component of the current density, $J_{z}(t)$, for the same magnetic field amplitude ($10^{5}$ Gauss) and different values of $\Delta$.

\section{Total power emission at electromagnetic spectra due to Synchroton radiation}
\label{AppendixB}

Following the production of charged particles due to the time-dependent background fields, these particles are subsequently accelerated by the strong background magnetic field. (A more detailed treatment of synchrotron radiation can be found in standard texts~\cite{Rybicki:847173,Pabmanabhan:2000book}.) 

To analyze the resulting electromagnetic radiation, we consider the Li\'enard-Wiechert potentials generated by a moving charge:
\begin{equation}
A_{0}(\vec{r}_{p},t) = \frac{e}{4\pi}\left(\frac{1}{r(1 - \hat{n}.\vec{\beta})}\right)_{\text{ret}}, 
\ \vec{A}(\vec{r}_{p},t) = \frac{e}{4\pi}\left(\frac{\vec{\beta}}{r(1 - \hat{n}.\vec{\beta})}\right)_{\text{ret}},
\end{equation} 
From these potentials, the electric and magnetic fields of the moving charge can be derived as:
\begin{equation}
\begin{split}
\vec{E}(t,\vec{r}) & = \frac{e}{4\pi\gamma^{2}}\left(\frac{(\hat{n} - \vec{\beta})}{r^{2}
(1 - \hat{n}.\vec{\beta})^{3}}\right)_{\text{ret}} + \frac{e}{4\pi}\left(\frac{\hat{n}\times
[(\hat{n} - \vec{\beta})\times\dot{\vec{\beta}}]}{r(1 - \hat{n}.\vec{\beta})^{3}}\right)_{\text{ret}}
\\
\vec{B}(t,\vec{r}) & = \hat{n}\times\vec{E}(t,\vec{r}),
\end{split}
\end{equation}
where $\beta = \vec{v}$ and $\gamma$ is the Lorentz factor. The power radiated by this accelerating charge is then given by the integral of the Poynting vector $\vec{S} = \vec{E}\times\vec{B}$ over a surface enclosing the charge:
\begin{equation}
P = \int \vec{S}.\hat{n}(1 - \hat{n}.\vec{\beta})r^{2}d\Omega.
\end{equation}
The power emission per unit solid angle is thus:
\begin{equation}
\frac{dP}{d\Omega} = \vec{E}^{2}(1 - \hat{n}.\vec{\beta})r^{2} = \frac{e^{2}}{(4\pi)^{2}}\frac{[\hat{n}\times[(\hat{n} - \vec{\beta})\times\dot{\vec{\beta}}]]^{2}}{(1 - \hat{n}.\vec{\beta})^{5}}.
\end{equation}
In the relativistic regime, the denominator of this expression is sharply peaked in the forward direction. Integrating the angular distribution yields the total radiated power:
\begin{equation}
P = \frac{e^{2}}{6\pi}\gamma^{6}[\dot{\vec{\beta}}^{2} - (\vec{\beta}\times\dot{\vec{\beta}})
^{2}].
\end{equation}
This can be decomposed into cases where the acceleration is parallel or perpendicular to the velocity. For parallel acceleration ($\vec{\beta} \times \dot{\vec{\beta}} = 0$):
\begin{equation}
P_{\parallel} = \frac{e^{2}}{6\pi} \gamma^{6}\dot{\vec{\beta}}_{\parallel}^{2} = \frac{e^{2}}
{6\pi m^{2}}\left(\frac{d\vec{p}}{dt}\right)^{2}.
\end{equation}
And for perpendicular acceleration ($\dot{\vec{\beta}}^{2} - (\vec{\beta}\times\dot{\vec{\beta}})^{2} = \dot{\vec{\beta}}_{\perp}^{2}/\gamma^{2}$):
\begin{equation}
P_{\perp} = \frac{e^{2}}{6\pi} \gamma^{4}\dot{\vec{\beta}}_{\perp}^{2} = \frac{e^{2}}{6\pi m^{2}}\gamma^{2}\left(\frac{d\vec{p}}{dt}\right)^{2}.
\end{equation}
These expressions highlight that for a given applied force $d\vec{p}/dt$, perpendicular acceleration results in $\gamma^{2}$ times more radiated power compared to parallel acceleration. In the specific case of circular motion with a bending radius $\rho$, the rate of change of momentum is:
\begin{equation}
\frac{d\vec{p}}{dt} = \frac{\gamma m\beta^{2}}{\rho}.
\end{equation}
Using this, we can now determine the instantaneous power radiated by a single electron of mass $m$ and energy $E$:
\begin{equation}
P_{\perp} = \frac{e^{2}}{6\pi m^{4}}\frac{E^{4}}{\rho^{2}}.
\end{equation}
Following the acceleration of the produced charged particles by the background magnetic field, they will emit synchrotron radiation. To determine the characteristics of this radiation, we begin by considering the Fourier transform of the time-varying electric field:
\begin{equation}
\vec{E}(\omega) = \frac{1}{\sqrt{2\pi}}\int_{-\infty}^{\infty}\vec{E}(t)e^{i\omega t}dt.
\end{equation}
The total energy received by an observer per unit solid angle over a single particle passage is then related to this Fourier transform:
\begin{equation}
\frac{dW}{d\Omega} = \int\frac{dP}{d\Omega}dt = \int_{-\infty}^{\infty}(r\vec{E}(t))^{2}dt
= \frac{1}{\pi}\int_{0}^{\infty}(r\vec{E}(\omega))^{2}d\omega.
\end{equation}
This leads to the angular and frequency distribution of the received energy:
\begin{equation}
\frac{d^{2}W}{d\Omega d\omega} = \frac{1}{\pi}\Big|\int_{-\infty}^{\infty}(r\vec{E}(t))e^{i\omega t}dt\Big|^{2}.
\end{equation}
In the far-field approximation, this expression becomes:
\begin{equation}
\frac{d^{2}W}{d\Omega d\omega} = \frac{e^{2}}{16\pi^{3}}\Big|\int_{-\infty}^{\infty}
\left(\frac{\hat{n}\times[(\hat{n} - \vec{\beta}) \times \dot{\vec{\beta}}]}{(1 - \hat{n}.\vec{\beta})^{3}}\right)_{\text{ret}}e^{i\omega t}dt\Big|^{2}.
\end{equation}
For particles moving in a circular path due to the magnetic field, with trajectory
\begin{equation}
\vec{R}(t) = \left(\rho\left(1 - \cos\frac{\beta t}{\rho}\right), 0, \rho\sin\frac{\beta t}
{\rho}\right),
\end{equation}
the relevant cross-product is
\begin{equation}
\hat{n}\times(\hat{n}\times\vec{\beta}) = \beta\left( - \sin\frac{\beta t}{\rho}\hat{z} +
\cos\frac{\beta t}{\rho}\sin\theta \hat{y}\right).
\end{equation}
Substituting this into the energy distribution and simplifying yields the standard synchrotron radiation spectrum:
\begin{equation}
\frac{d^{2}W}{d\Omega d\omega} = \frac{e^{2}}{16\pi^{3}}\gamma^{2}\left(\frac{\omega}{\omega_{c}}\right)^{2}(1 + \gamma^{2}\theta^{2})^{2}\Big[K_{2/3}^{2}(\xi) + \frac{\gamma^{2}\theta^{2}}{1 + \gamma^{2}\theta^{2}} K_{1/3}^{2}(\xi)\Big],
\end{equation}
where $\theta$ is the vertical observation angle and
\begin{equation}
\xi = \frac{\omega}{2\omega_{c}}(1 + \gamma^{2}\theta^{2})^{3/2}, \ \omega_{c} = \frac{3\gamma^{2}eB}{4\pi mc}.
\end{equation}
For an observer at a distance $d$ from a single ultra-relativistic electron, the spectral flux density in the plane of acceleration is given by:
\begin{equation}
S_{e}(\omega) = \frac{\sqrt{3}e^{3}B}{4\pi d^{2}mc^{2}}F\left(\frac{\omega}{\omega_{c}}\right),
\ F(x) = x\int_{x}^{\infty}K_{5/3}(\xi)d\xi.
\end{equation}

The derived synchrotron radiation spectrum provides a crucial link between the quantum particle production in the dynamic magnetic field and potentially observable electromagnetic emission. The characteristic frequency $\omega_c$ scales with the square of the Lorentz factor $\gamma^2$ and the magnetic field strength $B$, indicating that highly energetic particles in strong magnetic fields will radiate at higher frequencies. The spectral flux density $S_e(\omega)$ depends on the magnetic field strength and the distance to the observer, allowing for estimations of the expected signal strength from astrophysical sources where such particle production is significant. The specific form of the function $F(\omega/\omega_c)$ predicts a broad, non-thermal spectrum, distinct from blackbody radiation, which can serve as a diagnostic signature for synchrotron emission from these quantum-mechanically produced particles. This theoretical framework lays the groundwork for predicting and interpreting potential radio and higher-frequency observations of extreme astrophysical environments, such as the vicinity of black holes, where strong, time-varying magnetic fields are expected. The unique spectral characteristics and polarization properties of synchrotron radiation offer a powerful tool to indirectly probe the quantum processes occurring in these otherwise inaccessible regions of the universe.

%


\end{document}